\documentclass{article}

\usepackage{arxiv}

\usepackage[utf8]{inputenc} 
\usepackage[T1]{fontenc}    
\usepackage{hyperref}       
\usepackage{url}            
\usepackage{booktabs}       
\usepackage{amsfonts}       
\usepackage{nicefrac}       
\usepackage{microtype}      
\usepackage{cleveref}       
\usepackage{lipsum}         
\usepackage{graphicx}
\usepackage[square,numbers]{natbib}
\usepackage{doi}

\title{Improving the low-dose performance of aberration correction in single sideband ptychography}

\date{}

\newif\ifuniqueAffiliation
\uniqueAffiliationtrue

\usepackage{authblk}

\author[1,2,*]{{Songge Li}
}
\author[1,2]{{Nicolas Gauquelin}
}
\author[1,2]{{Hoelen L. Lalandec Robert}
}
\author[1,2]{{Arno Annys}
}
\author[3]{{Chuang Gao}
}
\author[1,2]{{Christoph Hofer}
}
\author[1,2]{{Timothy J. Pennycook}
}
\author[1,2]{{Jo Verbeeck}
}
\affil[1]{Electron Microscopy for Materials Science (EMAT), University of Antwerp, Groenenborgerlaan 171, 2020 Antwerp, Belgium}
\affil[2]{NANOlight Center of Excellence, University of Antwerp, Groenenborgerlaan 171, 2020 Antwerp, Belgium}
\affil[3]{Center of Analytical Facilities, Nanjing University of Science and Technology, Xiaolingwei Street 200, Nanjing, China}
\affil[*]{Corresponding author: songge.li@uantwerpen.be}

\renewcommand{\shorttitle}{Improving the low-dose performance of aberration correction in single sideband ptychography}

\begin{document}
This is the author's accepted manuscript (AAM), post peer review. The final published version is available at: \href{https://doi.org/10.1016/j.ultramic.2025.114225}{10.1016/j.ultramic.2025.114225}
\\
\\
\maketitle

\begin{abstract}
The single sideband (SSB) framework of analytical electron ptychography can account for the presence of residual geometrical aberrations induced by the probe-forming lens. However, the accuracy of this aberration correction method is highly sensitive to the invested electron dose, in part due to the necessity of phase unwrapping. In this work, we thus propose two strategies to improve the performance in low-dose conditions: confining phase unwrapping within the sidebands and selecting only well-unwrapped sidebands for calculating aberration coefficients. These strategies are validated through SSB reconstructions of both simulated and experimental 4D-STEM datasets of monolayer tungsten diselenide (WSe\textsubscript{2}). A comparison of results demonstrates significant improvements in Poisson noise tolerance, making aberration correction more robust and reliable for low-dose imaging.
\end{abstract}


\section{Introduction}

Scanning transmission electron microscopy (STEM) has become a powerful tool for exploring the atomic-scale structure of materials, owing to the short wavelength of electrons and the widespread adoption of aberration correctors\cite{Haider1998, Dellby2001, Batson2002}. In recent years, the development of fast pixelated detectors has enabled the acquisition of four-dimensional STEM (4D-STEM) datasets\cite{Plackett2013, Poikela2014, Tate2016, Ballabriga2011, Ryll2016, Llopart2022, Zambon2023a, Ercius2024}. Based on overlapping illuminated areas within the scan raster, electron ptychography permits the retrieval of the complex transmission function of specimens\cite{Rodenburg1992, Rodenburg1993, Nellist1995, Faulkner2004, Rodenburg2004, Guizar-Sicairos2008, Maiden2009}. This computational imaging technique is strongly valued for its high dose-efficiency\cite{Pennycook2015, Yang2015, Yang2015b, OLeary2020, OLeary2021, LalandecRobert2025}, its ability to achieve resolution beyond the diffraction limit\cite{Rodenburg1992, Nellist1995, Maiden2011, Humphry2012} and the possibility it offers to account for geometrical aberrations\cite{Maiden2009, Thibault2009, Yang2016a}. These advantages make ptychography a powerful tool for characterizing materials with sub-angstrom resolution\cite{Jiang2018}, enabling low-dose imaging for beam-sensitive materials\cite{Zhou2020b, Dong2023, Scheid2023, Hao2023, Li2025}, and understanding the physical properties of specimens\cite{Chen2022, Hofer2023b}.

Residual geometrical aberrations represent a major challenge in STEM imaging, as they introduce image distortions and may thus preclude the extraction of structural information. Due to its capacity to account for those in the reconstruction, ptychography has however made this less of a limiting factor\cite{Nguyen2024}, though an influence of strong aberrations can remain in the transfer of frequency information\cite{Koppell2021, Dwyer2024, LalandecRobert2025}. In particular, iterative ptychography methods, such as the extended ptychographic iterative engine (ePIE)\cite{Maiden2009}, have been widely adopted due to their ability to simultaneously reconstruct both the object and probe functions. However, the success of ePIE highly depends on the initialization of the algorithm and on the values attributed to the update strengths for both object and probe. Without appropriate choices for those parameters, the algorithm may fail to converge or become trapped at a local minimum instead of the correct solution. Additionally, iterative ptychography is often numerically intensive, thus sometimes requiring advanced computation capacities or long calculation time\cite{Yu2022, Wang2022a, Mukherjee2022, Welborn2024}.

In contrast, analytical ptychography methods, i.e. the Wigner distribution deconvolution (WDD)\cite{Rodenburg1992, Bates1989, Li2014} and the single sideband (SSB) approach\cite{Rodenburg1993, Pennycook2015}, provide mathematical frameworks for directly calculating the transmission function of specimens. They are favored for their robustness and capacity for live processing\cite{Strauch2021, Bangun2023}. However, during the reconstruction process, since the probe is conventionally assumed to be ideal, the accuracy of the result heavily relies on precisely corrected aberrations in experiments. To include aberrations under the weak phase object approximation (WPOA), the SSB framework offers a solution by flattening the sidebands with parameterized aberration coefficients\cite{Yang2016a}. While this was successfully demonstrated in experiments, challenges can be encountered when attempting to determine the aberrations present at low doses due to the difficulty of performing phase unwrapping with low signal-to-noise ratios (SNR).

In this paper, we propose two approaches for optimizing the low-dose performance of aberration correction within the SSB framework. The first involves exclusively unwrapping the phase within the sidebands, and the second consists of a careful selection strategy that employs only well-unwrapped sidebands for calculating the aberrations. We demonstrate the effectiveness of these methods using both simulated and experimental 4D-STEM datasets of monolayer tungsten diselenide (WSe\textsubscript{2}), illuminated using an acceleration voltage of 60 kV. In the simulations, a randomly aberrated probe is employed, while in the experiment, the probe was left in an imperfect correction state on purpose.
\section{Theoretical aspects}

\subsection{Single sideband ptychography}

\begin{figure}
    \centering
    \includegraphics[width=0.6\columnwidth]{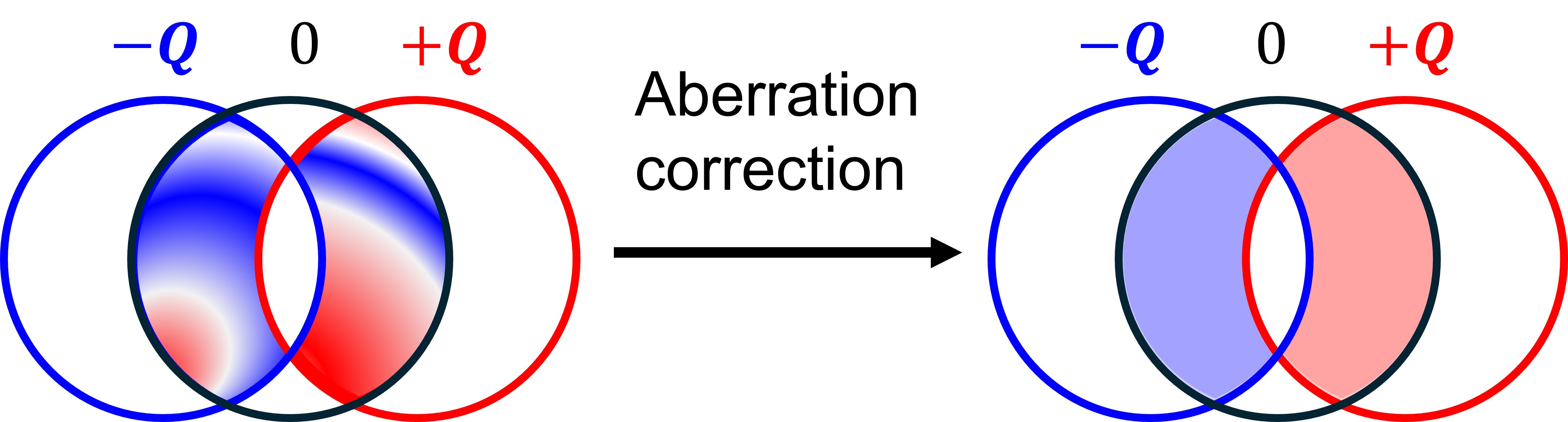}
    \caption{Aberration correction process in SSB ptychography. Schematic sidebands before (left) and after correction (right).}
    \label{fig_1}
\end{figure}

In STEM, following the interaction between the electron probe and the specimen, a convergent beam electron diffraction (CBED) pattern is formed in the far-field. As the probe is scanned across an imaged area, a 4D dataset, denoted as $I(\mathbf{K,R})$, can be generated using a pixelated electron detector. Here $\mathbf{K}$ is a scattering vector, and $\mathbf{R}$ is a scan position. By applying a Fourier transformation (FT) from position $\mathbf{R}$ to frequency $\mathbf{Q}$, this dataset is converted into a distribution $G(\mathbf{K,Q})$. Under the WPOA, $G(\mathbf{K,Q})$ can be simplified to:
\begin{equation}
    \label{eq:WPOA}
    \begin{aligned}
        G(\mathbf{K,Q}) = & |T(\mathbf{K})|^2 \delta(\mathbf{Q}) \\
        & + iT^* (\mathbf{K})T(\mathbf{K-Q}) \phi(\mathbf{Q}) \\
        & - iT(\mathbf{K})T^* (\mathbf{K+Q}) \phi^*(-\mathbf{Q})
    \end{aligned}
\end{equation}
Here, $T(\mathbf{K})=A(\mathbf{K})e^{i\chi(\mathbf{K})}$ represents the probe function in reciprocal space, where $A(\mathbf{K})$ is an aperture and $\chi(\mathbf{K})$ is the aberration function. $\phi(\mathbf{Q})$ represents the Fourier transform of the phase shift map. Since the probe is generated with a circular aperture, $T(\mathbf{K})$ possesses a circular shape in reciprocal space as well. Therefore, for a given $\mathbf{Q}$, Eq. \ref{eq:WPOA} can be visually represented by the intersection of three circles, with radii equal to the maximum frequency passing through the numerical aperture, and whose centers are located at frequencies $\mathbf{0}$, $-\mathbf{Q}$ and $\mathbf{Q}$, respectively, as shown in Fig. \ref{fig_1}. The function transferred in the blue disk is $- iT(\mathbf{K})T^* (\mathbf{K+Q}) \phi^*(-\mathbf{Q})$, and the one in the red disk is $iT^* (\mathbf{K})T(\mathbf{(K-Q}) \phi(\mathbf{Q})$. Ideally, when the probe is unaberrated, meaning that $T(\mathbf{K})==1$ in the circled area, the double overlapped regions, i.e. the sidebands, can be treated as flat, and thus homogeneously equal to $-i\phi^*(-\mathbf{Q})$ and $i\phi(\mathbf{Q})$. By extracting $\phi(\mathbf{Q})$ for all $\mathbf{Q}$ from the distribution $G(\mathbf{K,Q})$, and then performing an inverse FT to obtain $\phi(\mathbf{r})$, the SSB ptychographic reconstruction is completed. Note that the inverse FT of $\phi(\mathbf{Q})$ and $\phi^*(-\mathbf{Q})$ give the same result, as the phase shift $\phi(\mathbf{r})$ in real space is a real-numbered function. As a consequence, the last two terms of Eq. \ref{eq:WPOA} cancel out and no information is transferred in the triple overlapped region. 

\subsection{Aberration correction}

SSB provides a framework for calculating the aberration coefficients, by compensating the aberrated sidebands (left part of Fig. \ref{fig_1}) to flat (right part). In this correction process, the aberration function is separated into terms of several orders, with coefficients used as a collection of optimization parameters. In the aberrated sideband, the quantity ($\angle G(\mathbf{K,Q})$) is the superposition of the angle of $\phi(\mathbf{Q})$ and of the aberration function $\chi(\mathbf{K})$. For example, for the double overlap regions in the red disk, the angle of function $G(\mathbf{K,Q})$ is:
\begin{equation}
    \label{eq:G phase}
    \angle G(\mathbf{K,Q}) = \angle\phi(\mathbf{Q}) - \chi(\mathbf{K}) + \chi(\mathbf{K-Q})
\end{equation}
The remaining task is to solve the multiple equations from Eq. \ref{eq:G phase}. For instance, when considering only the first order, they can be described by three unknown parameters: $C_1$, $C_{12a}$ and $C_{12b}$\cite{Scherzer1936, Scherzer1947a, Sawada2008}. Additionally, the Fourier transformed phase shift map $\phi(\mathbf{Q})$ introduces a fourth unknown, leading to a requirement of at least four independent equations for a unique solution. To obtain these equations, we select four different points within the sideband, each corresponding to a specific value of $\mathbf{K}$, which in principle provides sufficient information to calculate the parameterized coefficients. To improve the accuracy and minimize computational errors, we compile all available equations from multiple selected sidebands, each corresponding to a different $\mathbf{Q}$, and solve them collectively using the singular value decomposition (SVD) method. Note that in the following aberration correction procedure, we further extend our consideration to higher-order aberration terms, including coefficients up to the fifth order.

\subsection{Phase jump problem at low dose}

A significant challenge appears when solving Eq. \ref{eq:G phase}: $\angle G(\mathbf{K,Q})$ must be continuous inside the sideband, while numerical angle values undergo discontinuous jumps from $\pi$ to $-\pi$ due to mathematical restrictions. This issue is evident in Fig. \ref{fig_2}(a). Here, we select the sideband with the highest frequency weighting, highlighted with a blue circle, as determined by the integral of the function $\int G(\mathbf{K,Q}) d\mathbf{K}$. In the selected sideband, there are some discontinuous areas where phase jumps from red ($\pi$) to blue ($-\pi$), which breaks continuity. Therefore, it is necessary to employ a phase unwrapping algorithm\cite{Herraez2002}, which however might fail due to the low SNR at low-dose conditions. We thus propose two new methods to alleviate this problem, which are described in the following section.

\section{Methodology}

\subsection{Confining phase unwrapping}

Here, a simulated 4D-STEM dataset of monolayer WSe\textsubscript{2} is employed as the basis for object reconstruction, given a randomly aberrated probe. The simulation parameters include an acceleration voltage of 60 kV and a semi-convergence angle of 25 mrad. The electron dose is limited to 2000 \(\mathrm{e^-}/\text{\AA}^2\) by applying Poisson noise to an initially noise-free simulated dataset (infinite dose). This low-dose simulation is employed to evaluate and compare the performance of the different phase unwrapping methods under noisy conditions.

\begin{figure}
    \centering
    \includegraphics[width=0.6\columnwidth]{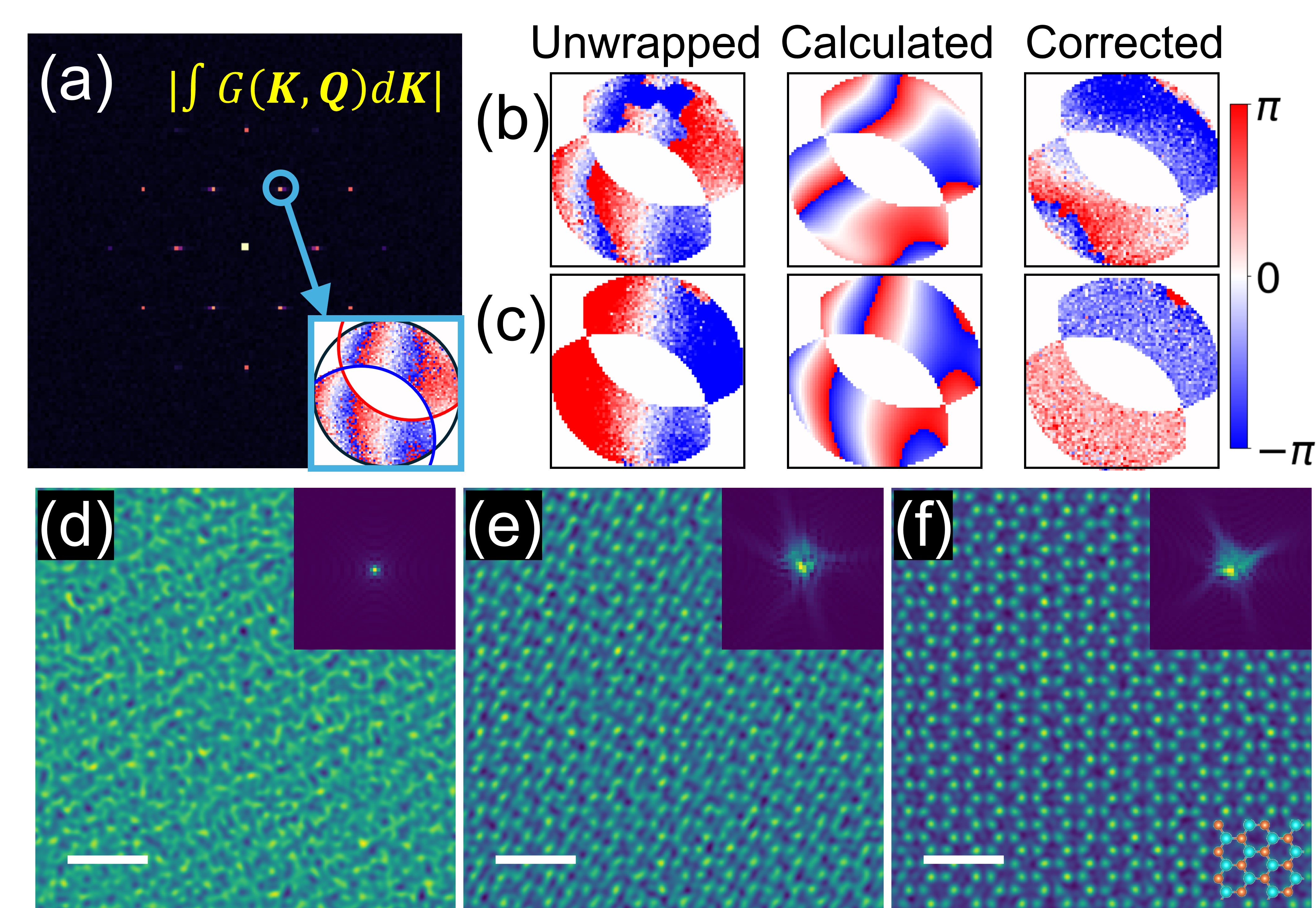}
    \caption{Aberration correction by two different phase unwrapping approaches for a simulated dataset of monolayer WSe\textsubscript{2}, with an accelerating voltage of 60 kV, a semi-convergence angle of 25 mrad, a randomly generated probe, and an electron dose of 2000 \(\mathrm{e^-}/\text{\AA}^2\). (a) Amplitude value of integral \( \int G(\mathbf{K}, \mathbf{Q}) \, d\mathbf{K} \) as a function of frequency \(\mathbf{Q}\). The sideband with the circled index is selected to illustrate the aberration correction process. (b) Unconfined phase unwrapping approach: unwrap phase across all coordinates $\mathbf{K}$ for a fixed $\mathbf{Q}$ then set the value outside the sidebands to 0, then correct. (c) Confined approach: unwrap phase exclusively within the sidebands, leading to their flattening. (d,e,f) SSB reconstructions with employed probe amplitude shown as insets. The process is performed (d) with no aberration correction, (e) with aberration correction using unconfined phase unwrapping method and (f) with unwrapping exclusively within the sideband. A top-view model of monolayer WSe\textsubscript{2} is included in (f), where cyan atoms represent W and the orange represent Se\textsubscript{2}. Scale bar: 1 nm.}
\label{fig_2}
\end{figure}

The first approach tested involves unwrapping the phase across all coordinates $\mathbf{K}$ for a given $\mathbf{Q}$ and then setting the values outside the sidebands to zero. The result of applying this process to the wrapped sideband in Fig. \ref{fig_2}(a) is shown in the left image of Fig. \ref{fig_2}(b), where unwrapping remains imperfect. Based on this, we calculate the aberration coefficients, using the approach described in Ref.\cite{Yang2016a}, and use this calculated sideband (the middle image of Fig. \ref{fig_2}(b)) to compensate the original. This leads to the corrected sideband shown in the right image of Fig. \ref{fig_2}(b). Note that the aberration correction extends up to fifth order, whereas the originally published method\cite{Yang2016a} only went as high as third order. However, as demonstrated, the final corrected sideband remains non-flat. 

In the second approach, phase unwrapping is confined within the sideband, without considering the areas outside. This successfully produces a well-unwrapped, continuous, phase, as shown in the left image of Fig. \ref{fig_2}(c). Using the unwrapped phase, we calculate the aberration coefficients and use them to correct the original phase. The final result, shown in the right image of Fig. \ref{fig_2}(c), demonstrates a much flatter corrected phase compared to the first method. 

We compare the uncorrected SSB reconstruction and the two corrected cases. Fig. \ref{fig_2}(d) shows the phase image of the direct SSB reconstruction obtained without aberration correction, assuming a perfect probe. As expected, it shows no discernible WSe\textsubscript{2} crystal information due to aberration-induced distortions. Fig. \ref{fig_2}(e) shows the reconstructed image and the calculated probe amplitude based on the unconfined phase unwrapping approach, which reveals some crystal information but still leads to some clear distortions owing to the remaining aberrations. Finally, Fig. \ref{fig_2}(f) shows the reconstructed image and the calculated probe based on the confined phase unwrapping approach. Here, the aberration is effectively corrected, resulting in a high-quality reconstruction with clear atomic details, which appropriately matches with the monolayer WSe\textsubscript{2} model. These results demonstrate the critical importance of a robust unwrapping process, and highlight the confined method as a more reliable approach for aberration correction in SSB reconstruction.

\subsection{Employing only well-unwrapped sidebands}

Even with the most advanced phase unwrapping methods, a reduction of the dose below e.g. 1000 \(\mathrm{e^-}/\text{\AA}^2\), as shown in our simulation, remains challenging for aberration correction due to the extremely low SNR. We thus propose a sideband selection strategy to complete the phase unwrapping method. To obtain the highest SNR, it seems natural to choose the sidebands having the highest frequency weighting, estimated through $|\int G(\mathbf{K,Q}) d\mathbf{K}|$. To improve the precision of the aberration calculation process, a user-defined number of sidebands is thus employed to extract the coefficients. These high-intensity sidebands are typically expected to possess a higher SNR, making them more reliable for the phase unwrapping process. In addition, since the confined phase unwrapping method has demonstrated superior performance, it is employed as the only phase unwrapping method in the following.

\begin{figure}
    \centering
    \includegraphics[width=0.6\columnwidth]{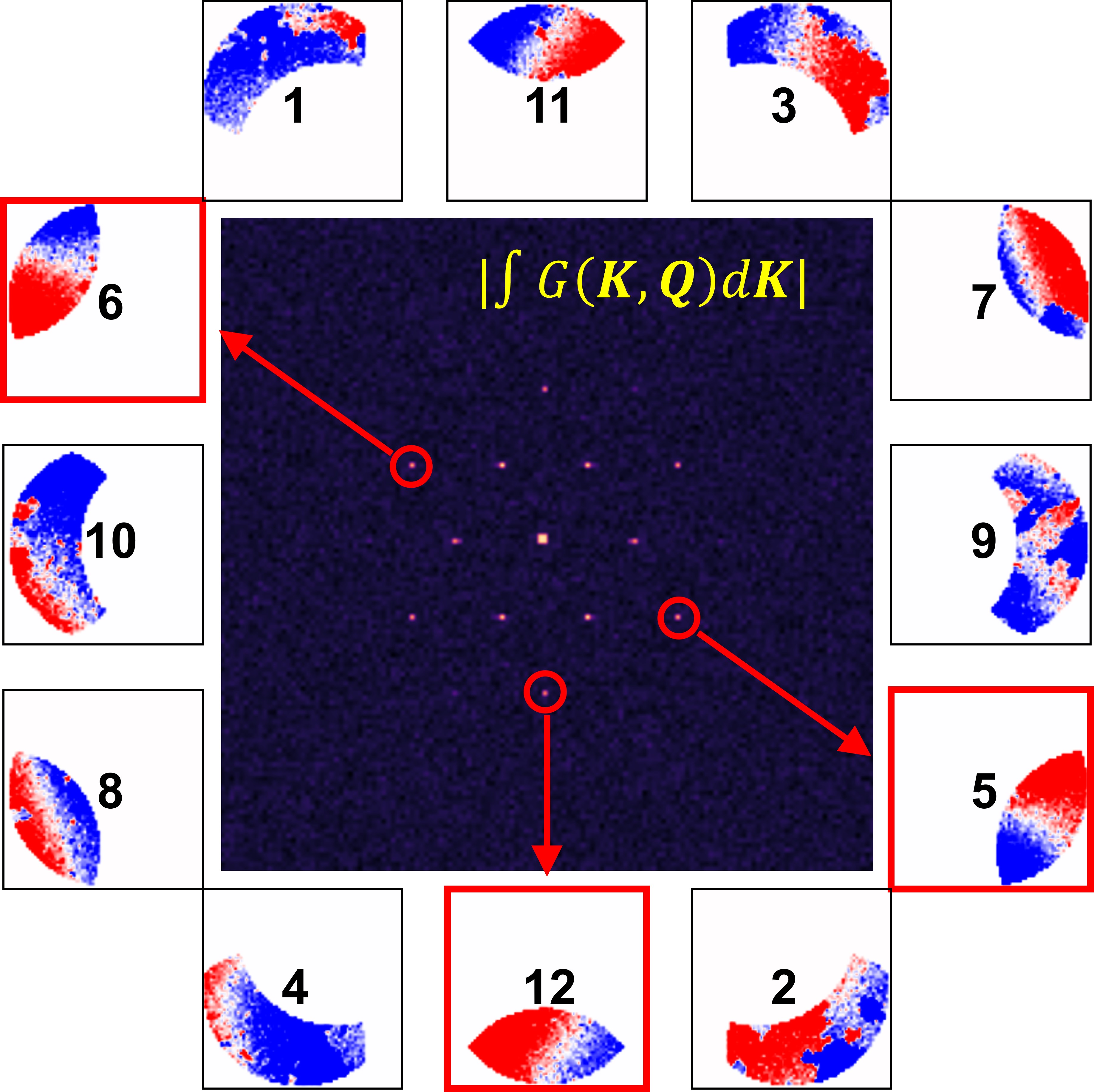}
    \caption{Unwrapped phase of the right sidebands for different frequencies $\mathbf{Q}$ for the dataset with a reduced dose of 500 \(\mathrm{e^-}/\text{\AA}^2\). The indices are denoted according to the frequency weighting \( |\int G(\mathbf{K}, \mathbf{Q}) \, d\mathbf{K}| \) of the sidebands, arranged from high to low. Only the sidebands with indices 5, 6 and 12 exhibit the successful unwrapping.}
\label{fig_3}
\end{figure}

Though counterintuitive at first sight, we find that a lower noise level does not necessarily correspond to an easier unwrapping process. To illustrate this, we show results for 12 sidebands in Fig. \ref{fig_3}. Here, the same simulated dataset but with a lower dose of 500 \(\mathrm{e^-}/\text{\AA}^2\) is employed. The sidebands are arranged by frequency weighting, with index 1 corresponding to the highest and index 12 to the lowest. We note that not all high-intensity sidebands are perfectly well-unwrapped. Only the sidebands with indices 5, 6 and 12 exhibit successful unwrapping, while others fail. Note that for each frequency, only the right-leaning single sideband is selected, as the phase-unwrapping results are slightly asymmetric in such low-dose conditions, i.e. when comparing the left and right sidebands.

\begin{figure}
    \centering
    \includegraphics[width=0.6\columnwidth]{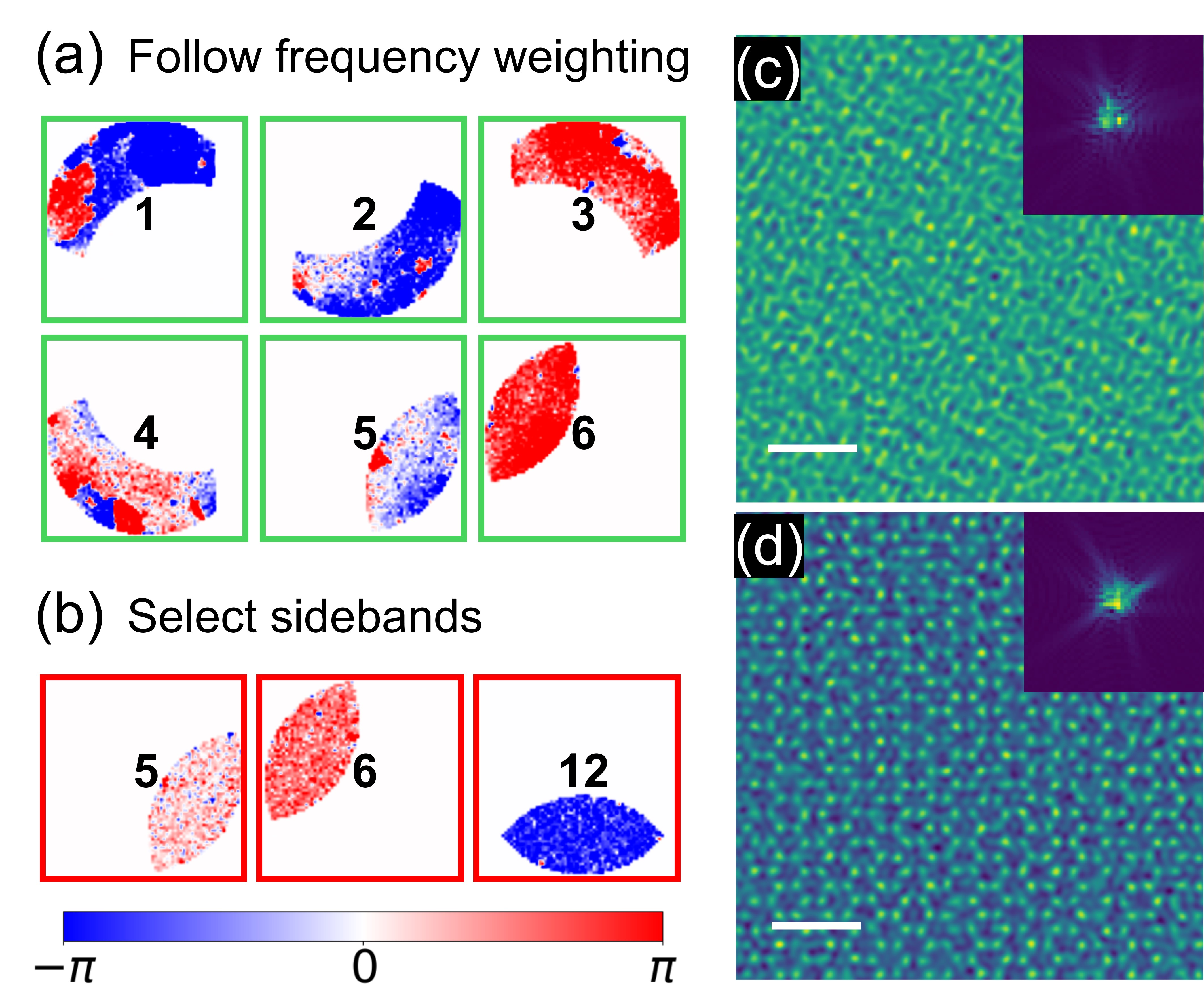}
    \caption{(a) Correction based on using the six highest-weighting sidebands (indices 1-6). (b) Improved aberration correction by selecting well-unwrapped (indices 5, 6 and 12) sidebands. (c, d) Reconstruction and calculated probe amplitude based on sideband selection in (a) and (b), respectively. Scale bar: 1 nm.}
\label{fig_4}
\end{figure}

To highlight the effect of the sideband selection strategy, the aberration coefficients are first calculated using the six highest-intensity cases (indices 1–6). The corrected sidebands in Fig. \ref{fig_4}(a) remain non-flat, indicating an imperfect correction. The reconstructed image and the calculated probe amplitude, obtained with these calculated coefficients, are shown in Fig. \ref{fig_4}(c). Effects from residual aberrations are clearly visible, degrading the quality of the reconstruction. In contrast, selecting only the well-unwrapped sidebands (indices 5, 6, and 12) results in significantly flatter corrected sidebands, as shown in Fig. \ref{fig_4}(b). The corresponding reconstruction shown in Fig. \ref{fig_4}(d) achieves clear atomic resolution and the calculated probe closely matches that in Fig. \ref{fig_2}(f), validating the effectiveness at a lower dose. These results thus confirm that this specific processing strategy, provided that the choice of sidebands is optimized, can enhance the performance of aberration correction at low electron doses.

\subsection{Comparison of performances}

\begin{figure*}
    \centering
    \includegraphics[width=1\textwidth]{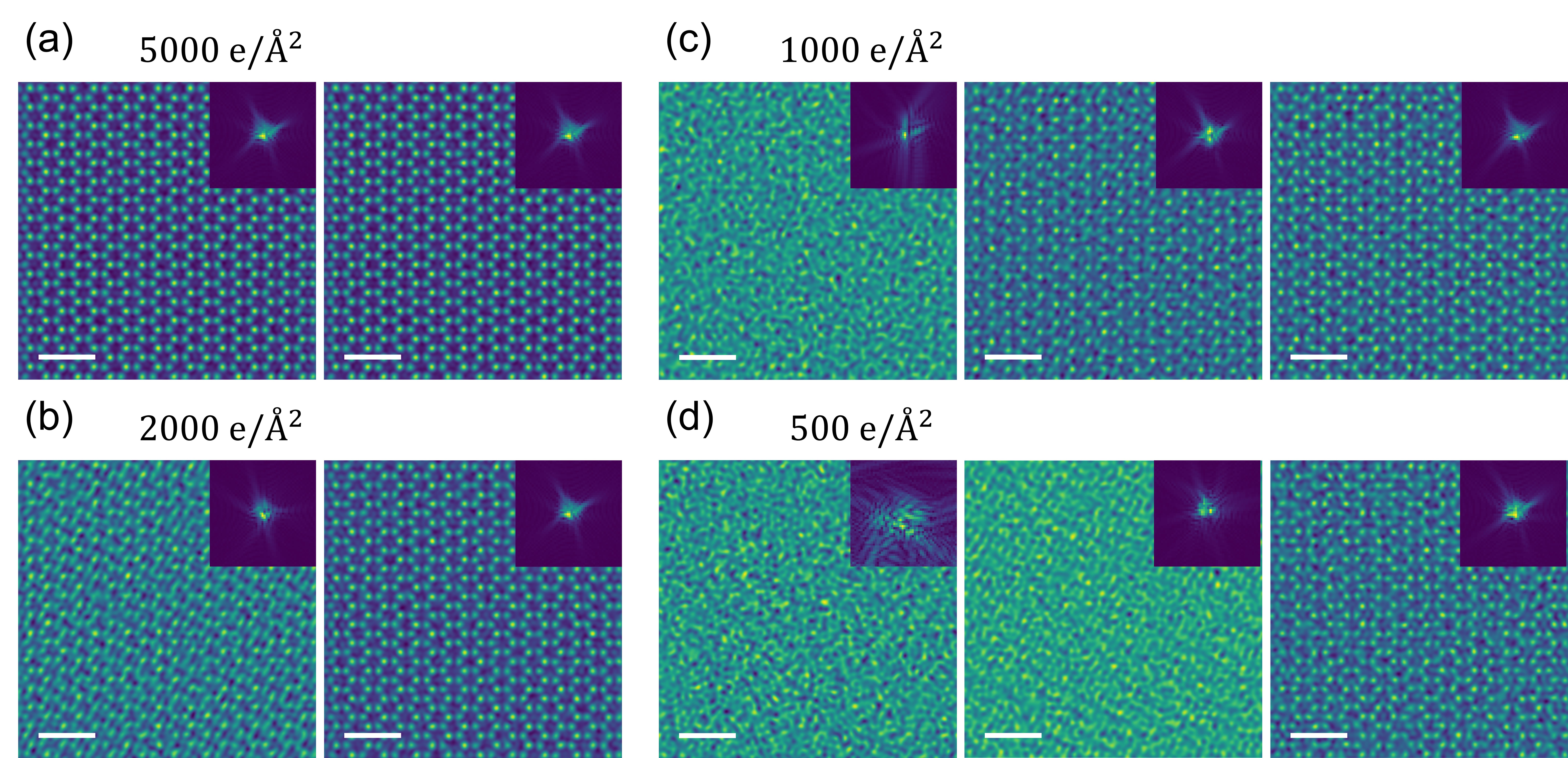}
    \caption{Comparison of different aberration correction approaches for dose series reconstruction, given simulated WSe\textsubscript{2} datasets. The left images of (a)-(d) show reconstructions using the unconfined phase unwrapping approach. The right images in (a, b) and the middle in (c, d) show reconstructions using the confined phase unwrapping approach and following the frequency weighting-based selection. The right images in (c, d) employ only well-unwrapped sidebands and resulting in improved reconstruction. Scale bar: 1 nm. }
\label{fig_5}
\end{figure*}

To evaluate the effectiveness of the two proposed approaches for low-dose imaging, we compare the SSB reconstructions before and after using them in Fig. \ref{fig_5}. As the sideband selection serves as an additional refinement to phase unwrapping, it is only applied when the confined phase unwrapping alone is insufficient. The left images of Fig. \ref{fig_5}(a)-(d) show the corrected reconstructions using the unconfined phase unwrapping method. There, aberrations are only well-corrected at 5000 \(\mathrm{e^-}/\text{\AA}^2\). At doses below 1000 \(\mathrm{e^-}/\text{\AA}^2\), the method fails completely, leading to a reconstruction with no useful crystal information. The right images in Fig. \ref{fig_5}(a) and (b) and the middle images in Fig. \ref{fig_5}(c) and (d) show the corrected reconstruction using the confined phase unwrapping method and following the frequency weighting-based selection strategy. This approach significantly improves the dose-efficiency of aberration correction, enabling improved reconstruction at low doses. However at the even lower doses used in Fig. \ref{fig_5}(c) and (d), significant residual aberrations remain, especially at 500 \(\mathrm{e^-}/\text{\AA}^2\) where the effect significantly degrades image quality. To address this, the optimized sideband selection strategy is applied to allow only the well-unwrapped sidebands to contribute to aberration calculation. The resulting reconstructions are shown in the right images in Fig. \ref{fig_5}(c) and (d), exhibiting further improvement.

\section{Experimental validation}

\begin{figure}
    \centering
    \includegraphics[width=0.6\columnwidth]{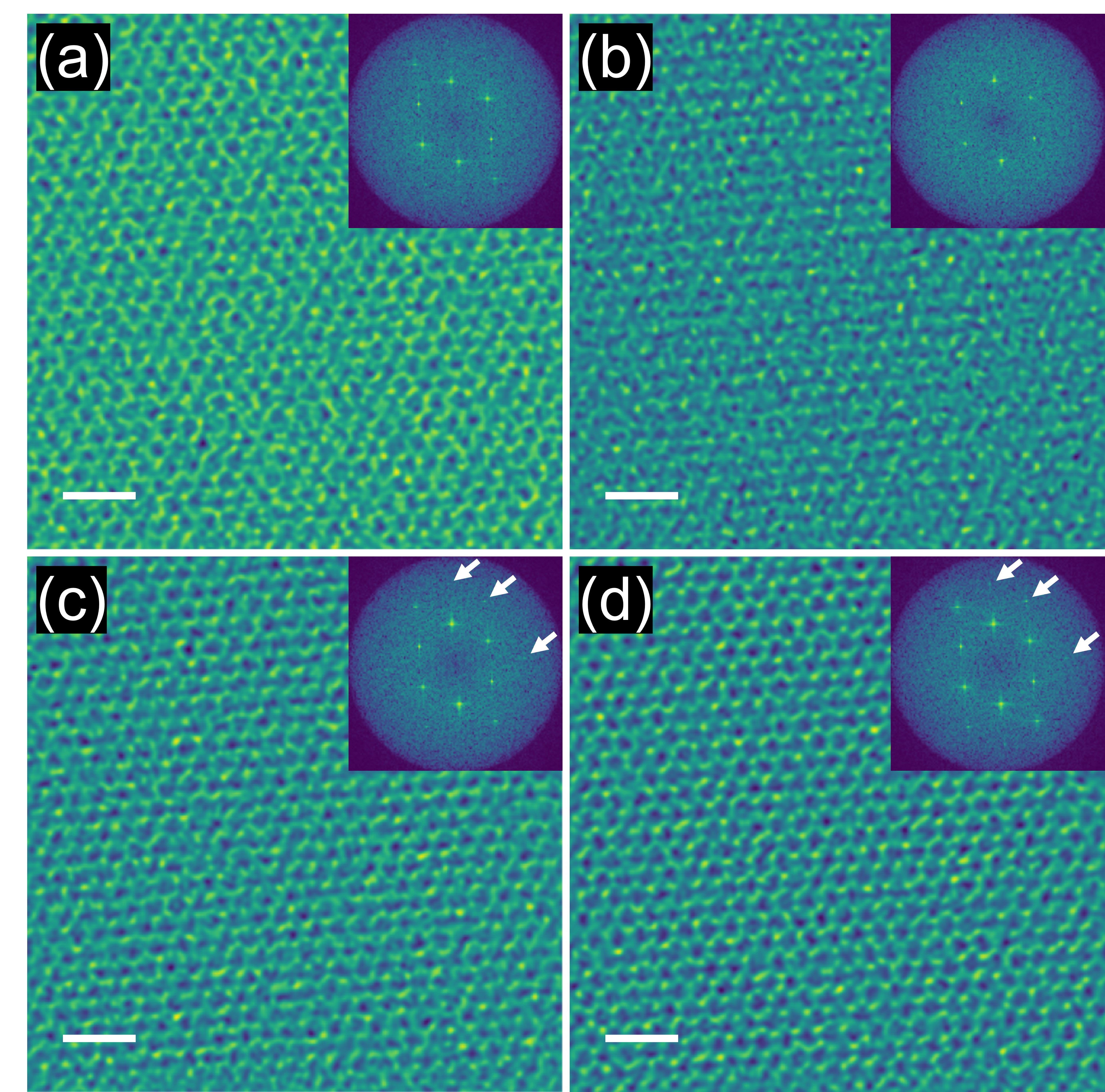}
    \caption{SSB reconstruction for experimental monolayer WSe\textsubscript{2} datasets, given an acceleration voltage of 60 kV, a semi-convergence angle of 25 mrad and a dose of 606 \(\mathrm{e^-}/\text{\AA}^2\). (a) Direct SSB reconstruction and its FT patterns. (b) Corrected reconstruction using the unconfined phase unwrapping method. (c) Corrected reconstruction using the confined phase unwrapping method. (d) Corrected reconstruction using both the confined phase unwrapping method and the optimized sideband selection strategy. White arrows in the FT patterns indicate spots that appear and enhance in (d) compared to (c). Scale bar: 1 nm.}
\label{fig_6}
\end{figure}

To experimentally validate the effectiveness of our proposed methods, we performed a 4D-STEM acquisition on a monolayer WSe\textsubscript{2}, using a Thermo Fisher Themis Z microscope equipped with a Timepix3 direct electron detector\cite{Poikela2014, Jannis2022a}. The measurement was performed with an acceleration voltage of 60 kV, a semi-convergence angle of 25 mrad and an electron dose of 606 \(\mathrm{e^-}/\text{\AA}^2\). Here, the electron dose is estimated by the method described in Ref.\cite{Jannis2022a}. To illustrate the post-acquisition correction process, residual aberrations were introduced during the experiment, resulting in noticeable degradation in the direct reconstruction and its FT patterns, as shown in Fig. \ref{fig_6}(a). The unconfined phase unwrapping method (Fig. \ref{fig_6}(b)) is applied but fails to correct the aberrations, leading to an even worse result. In contrast, the confined phase unwrapping method leads to a significantly improved reconstruction (Fig. \ref{fig_6}(c)), where the atoms are more visible than that in Fig. \ref{fig_6}(a). To achieve further refinement, the strategy of selecting only well-unwrapped sidebands is employed and yields an improved image in Fig. \ref{fig_6}(d). This is evidenced by the higher SNR of the FT pattern, particularly at the enhanced and the emerged spots indicated by white arrows. These experimental results confirm that our approaches can effectively improve aberration correction performance at low doses and demonstrate its practical applicability in experiments.

\section{Conclusion}

In conclusion, we report two combinable optimization approaches for aberration correction at low-dose condition, based on the SSB framework of analytical ptychography. Our findings illustrate that confining the phase unwrapping within the sideband and selecting only the well-unwrapped sidebands significantly improve the low-dose performance of this phase retrieval method. Tuning the microscope aberration corrector can be considerably more difficult at low doses, and therefore these advances can improve the performance of this fast, computationally effective and reliable analytical ptychography method for imaging beam-sensitive materials.

\section*{Author declarations}

\subsection*{Conflicts of interest}

The authors declare no conflicts of interest.

\subsection*{Author contribution}

\textbf{S.L.}: project conception, experiments, software, simulation, data treatment, results analysis, manuscript writing. \textbf{N.G.}: experiments. \textbf{H.L.L.R.}: results analysis, supervision, manuscript editing. \textbf{A.A.}: experiments. \textbf{C.G.}: data treatment. \textbf{C.H.}: data treatment. \textbf{T.J.P.}: supervision, manuscript editing. \textbf{J.V.}: supervision, manuscript editing.

\subsection*{Acknowledgments}

We thank Dr. Gang Wang from the Southern University of Science and Technology (Shenzhen, China) for kindly providing the monolayer WSe\textsubscript{2} TEM sample.

This work was funded by the FWO (Flanders, Belgium), under grants No. G042920N (Coincident event detection for advanced spectroscopy in transmission electron microscopy) and No. G013122N (Advancing 4D STEM for atomic scale structure property correlation in 2D materials). Further funding was received from the Horizon 2020 research and innovation programme (European Union), under grant agreement No. 101017720 (FET-Proactive EBEAM), as well as from the Horizon Europe framework program for research and innovation (European Union), under grant agreement No. 101094299 (IMPRESS). Views and opinions expressed are however those of the authors only and do not necessarily reflect those of the European Union or the European Research Executive Agency (REA). Neither the European Union nor the granting authority can be held responsible for them.

\subsection*{Data availability statement}

The source code and demonstration based on the original data that support the findings of this study are openly available in Zenodo at: https://doi.org/10.5281/zenodo.15309822

\bibliographystyle{unsrtnat}
\bibliography{SSBAC_paper}

\end{document}